\documentclass[aps,prb,twocolumn,showpacs,amsmath]{revtex4}
\usepackage{graphicx}
\usepackage{bm}

\begin{document}

\title{Josephson junctions in thin and narrow rectangular superconducting strips}

\author{John R.\ Clem}
\affiliation{%
   Ames Laboratory and Department of Physics and Astronomy, \\
   Iowa State University, Ames, Iowa, 50011--3160}

\date{\today}

\begin{abstract} 
I consider a Josephson junction crossing the middle of a thin rectangular superconducting strip of length $L$ and width $W$ subjected to a perpendicular magnetic induction $B$.  I calculate the spatial dependence of the gauge-invariant phase difference across the junction and the resulting $B$ dependence of the critical current $I_c(B)$.  
\end{abstract}


\maketitle

\section{\label{intro}Introduction}

Grain-boundary Josephson junctions play an important role in thin films of YBa$_2$Cu$_3$O$_{7-\delta}$ (YBCO).\cite{VanHarlingen95,Delin96,Hilgenkamp02,Durrell09}  Various theoretical approaches have been taken to understand the physics of Josephson junctions in thin films.\cite{Mints94,Kogan01,Moshe08}  When the film thickness $d$ is less than the London penetration depth $\lambda$, the current density $\bm j$ is practically uniform across the thickness, and the characteristic length governing the spatial distribution of the magnetic field distribution is the Pearl length,\cite{Pearl64} 
\begin{equation}
\Lambda = 2 \lambda^2/d.
\label{Pearl}
\end{equation}
However, various studies have shown that there is a nonlocal relationship between the Josephson-current distribution in the vicinity of a Josephson vortex core and the magnetic field these currents generate,\cite{Ivanchenko90,Gurevich92,Ivanchenko95,Kuzovlev97}  and the characteristic length describing the spatial variation of the gauge-invariant phase across the junction is (in SI units)
\begin{equation}
\ell = \phi_0/4\pi \mu_0 \lambda^2 j_c,
\label{ell}
\end{equation}
where $\phi_0 = h/2e$ is the superconducting flux quantum and $j_c$ (assumed to be independent of position) is the maximum Josephson current density that can flow as a supercurrent through the junction.

The integral equations relating the gauge-invariant phase difference across the junction to the magnetic field generated by a vortex in a long Josephson junction in a thin ($d < \lambda$) film of lateral dimensions large by comparison with $\Lambda$ and $\ell$ were examined analytically and solved numerically in Ref.\ \onlinecite{Kogan01} for arbitrary ratios of $\ell/\Lambda$.  The case of a short Josephson junction bisecting a long superconducting strip of width $W$ was studied in Ref.\ \onlinecite{Moshe08} under the assumptions that $W \ll \Lambda$ and $W \ll \ell$.  

\begin{figure}
\includegraphics[width=8cm]{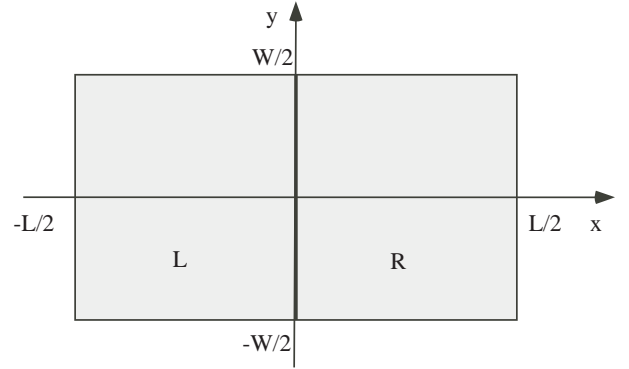}
\caption{%
Sketch of the thin rectangular superconducting strip of width $W$ and length $L$ considered here.  An in-line Josephson junction of width $d_i$  (bold line) is at the center at $x = 0$.  A magnetic induction $B$ is applied in the $z$ direction.  Current leads of separation $L$ (not shown) symmetrically feed current $I$ to the sample along the $x$ direction. }
\label{sample}
\end{figure}

In this paper I revisit the latter problem by considering a thin rectangular uniform superconducting strip of length $L$, width $W$, and thickness $d$ ($d < \lambda$) divided into two halves by a Josephson junction at $x = 0$, as shown in Fig.\ \ref{sample}. An applied magnetic induction $\bm B = \hat z B$  induces screening currents in the film.  However, I consider here only the simplest case for which $\Lambda = 2 \lambda^2/d$ is much larger than the smaller of $L$ and $W$, such that the self-field generated by the screening currents can be neglected.\cite{Moshe08}  
The purpose of this paper is to calculate how the screening currents induced in response to $\bm B$ affect the $B$ dependence of the maximum Josephson critical current $I_c(B)$.  

In Sec.\ II I give a brief discussion of the derivation of the basic equation for the gauge-invariant phase difference $\Delta \gamma(y)$ across the junction, in Sec.\ III I present the solutions  for $\Delta \gamma(y)$ for arbitrary ratios of $L/W$, and  in Sec.\ IV I briefly summarize the results.

\section{\label{phase}Gauge-invariant phase difference}

In the context of the Ginzburg-Landau (GL) theory,\cite{deGennes,StJames69} the superconducting order parameter can be expressed as $\Psi = \Psi_0 f e^{i\gamma}$, where $\Psi_0$ is the magnitude of the order parameter in equilibrium, $f = |\Psi|/\Psi_0$ is the reduced order parameter, and $\gamma$ is the phase.  The second GL equation (in SI units) is 
\begin{equation}
\bm j = -\frac{f^2}{\mu_0 \lambda^2}(\bm A +\frac{\phi_0}{2\pi}\nabla \gamma),
\label{GL2}
\end{equation} 
where $\bm A$ is the vector potential and  $\bm B = \nabla \times \bm A$ is the magnetic induction.  Since $\bm j$ is a gauge-invariant quantity, so is the quantity within the parentheses on the right-hand side. Different choices for the gauge of the vector potential $\bm A$ result in different expressions for $\gamma$.  

With a sinusoidal current-phase relation, the Josephson current density in the $x$ direction across the junction of width $d_i$ at $x = 0$ is $j_x(y) = j_c \sin \Delta \gamma (y)$, where $j_c$  is the maximum Josephson current density and $\Delta \gamma(y)$ is the gauge-invariant phase difference between the left (a) and right (b) superconductors,
\begin{equation}
\Delta \gamma(y) = \gamma_a(-\frac{d_i}{2},y) - \gamma_b(\frac{d_i}{2},y) -\frac{2\pi}{\phi_0}\int_{-d_i/2}^{d_i/2}
A_x(x,y) dx.  
\label{Deltagammadefinition}
\end{equation}

I assume here that the induced or applied current densities $\bm j_{\rm a}$ and $\bm j_{\rm b}$ on the left- and right-hand sides of the junction are so weak that the suppression of the magnitude of the superconducting order parameter  is negligible, such that $f = 1$.
A simple relation between these current densities and the gauge-invariant phase difference can be obtained by integrating the vector potential $\bm A$ around a very narrow rectangular loop of width $d_i$  in the $xy$ plane that just encloses the junction (with the bottom end at the origin and the top end at $y$), neglecting the magnetic flux up through the contour, making use of Eq.\ (\ref{GL2}) with $f = 1$ for those portions of the integration along the sides of the junction, and noting that, by symmetry,  $j_{{\rm a}y}(0,y) = - j_{{\rm b}y}(0,y)$:
\begin{equation}
\Delta \gamma(y) = \Delta \gamma_0+\frac{4\pi \mu_0 \lambda^2}{\phi_0}
\int_0^y j_{{\rm b}y}(0,y')dy',
\label{Deltagammaintegral}
\end{equation} 
where $ \Delta \gamma_0 = \Delta \gamma(0)$, such that 
\begin{equation}
d\Delta \gamma(y)/dy = (4\pi \mu_0 \lambda^2/\phi_0)
j_{{\rm b}y}(0,y).
\label{Deltagammaderivative}
\end{equation}

\section{\label{solution}Solving for $\Delta \gamma$}

I next assume that the Josephson coupling is so weak that the currents $\bm                                                                                j_{\rm a}$ and $\bm j_{\rm b}$  on the left- and right-hand sides of the junction induced in response to the applied magnetic induction $\bm B$ are far larger than the Josephson current density.  This is equivalent to the assumption that $W \ll \ell.$ Since  $\bm j_{\rm a}$ easily can be obtained by symmetry from $\bm j_{\rm b}$, I  calculate only $\bm j_{\rm b}$ in the region $x >0$ and suppress the subscript b.  

With the gauge choice $\bm A = -\hat x B y$, since $\nabla \cdot \bm j = 0$ [see Eq.\ (\ref{GL2})],  $\nabla^2 \gamma = 0$ must be solved subject to the boundary conditions following from  $j_x(0,y) = j_x(L/2,y)=0$ and  $j_y(x,\pm W/2) = 0$,
\begin{eqnarray}
&&\gamma_x(0,y)= \gamma_x(L/2,y)=2\pi B y/\phi_0,\\
&&\gamma_y(x, \pm W/2) = 0,
\end{eqnarray}
 where $\gamma_x = \partial \gamma/\partial x$ and $\gamma_y = \partial \gamma/\partial y$.
The solution, obtained by the method of separation of variables, is (up to a constant)
\begin{equation}
\gamma(x,y) = \frac{8\pi B}{\phi_0 W}\sum_{n=0}^\infty\frac{(-1)^n \sinh[k_n(x-L/4)]\sin (k_ny) }{k_n^3 \cosh(k_nL/4)},
\end{equation}
where $k_n = (n+1/2)2\pi/W$.  

\begin{figure}
\includegraphics[width=8cm]{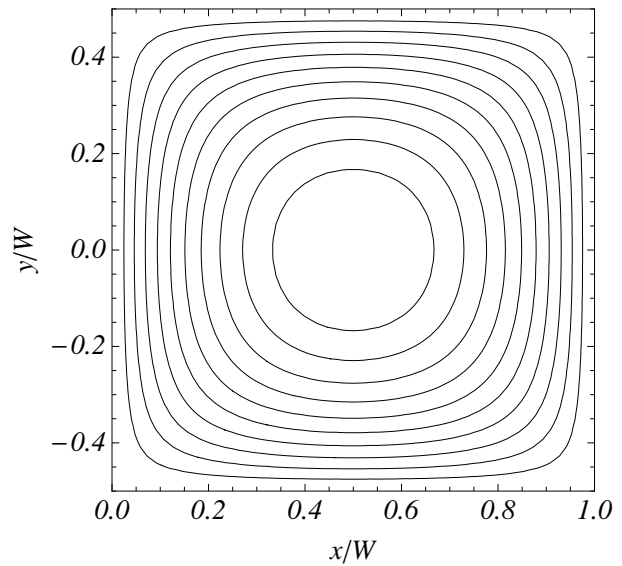}
\caption{%
Contour plot of the stream function $S(x,y)$ in the right half ($x > 0$) of a strip of length $L = 2W$.  The contours correspond to the streamlines of the current density $\bm j$ induced by the applied magnetic induction $\bm B = \hat z B$.}
\label{streamfunction}
\end{figure}

The current density $\bm j(x,y)$ now can be obtained from Eq.\ (\ref{GL2}).  Since $\nabla \cdot \bm j = 0$, we also can write $\bm j = \nabla \times \bm S$, where $\bm S = \hat z S$, and $S(x,y) = (B/2\mu_0 \lambda^2)s(x,y)$ is the stream function given by
\begin{equation}
s(x,y) = y^2 + \frac{8}{W}\sum_{n=0}^\infty \frac{(-1)^n \cosh[k_n(x-L/4)]\cos (k_ny) }{k_n^3 \cosh(k_nL/4)}.
\end{equation}
The series converges rapidly, and Fig.\ \ref{streamfunction} shows the result for $x>0$  obtained by summing over $n$ from 0 to 10 when $L/W = 2$.  

Since Eq.\ (\ref{GL2}) yields $j_y(0,y) = -(\phi_0/2\pi \mu_0 \lambda^2) \gamma_y(0,y)$, Eq.\ (\ref{Deltagammaderivative}) can be integrated to obtain (assuming the additive constant $\Delta \gamma_0 =0$)
\begin{equation}
\Delta \gamma(y) = \frac{16 \pi B}{\phi_0 W}\sum_{n=0}^\infty\frac{(-1)^n }{k_n^3}\tanh(k_nL/4) \sin (k_n y).
\label{Deltagamma}
\end{equation}
As shown in the Appendix, when $L \to \infty$, the sum can be expressed in terms of the Lerch transcendent $\Phi(z,s,a)$, and in this limit  $\Delta \gamma(y)$ also can be expressed as $\Delta \gamma(y) = (4BW^2/\phi_0)\varphi_0(\pi y/W)$, where $\varphi_0(\mu)$ is a function defined in Ref.\ \onlinecite{Moshe08}.  This is the explanation of why $\varphi_0(\mu)$ was found to be a  material-independent universal function\cite{Moshe08} when $\Lambda \gg W$ and $\ell \gg W$.

\begin{figure}
\includegraphics[width=8cm]{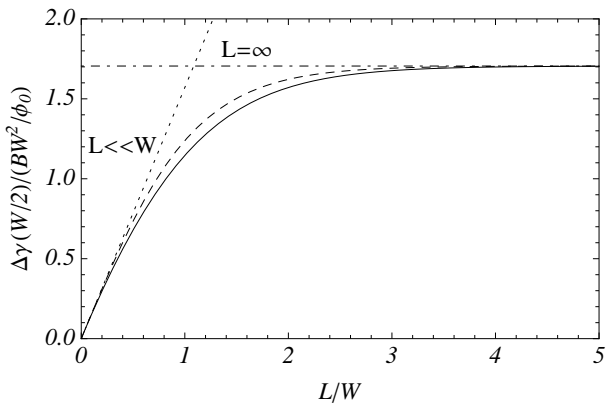}
\caption{%
$\Delta \gamma (W/2)$ (solid) vs $L/W$ and the approximations of Eqs.\ (\ref{Lbig}) (dot-dashed), (\ref{Lsmall}) (dotted), and (\ref{edgeapprox}) (dashed).}
\label{Deltagammaedgeplot}
\end{figure}

For all ratios of $L/W$, the maximum value of $\Delta \gamma(y)$ occurs at $y = W/2$, where
\begin{eqnarray}
\Delta \gamma(W/2)\!\! &=& \!\!\frac{14 \zeta(3) B W^2}{\pi^2 \phi_0}\! =\! 1.705 \frac{B W^2}{\phi_0},\; L \to \infty,
\label{Lbig}\\
\Delta \gamma(W/2)\!\! &=& \!\!\frac{\pi B W L}{2 \phi_0}\! = \!1.571 \frac{B W L}{\phi_0},\; L \ll W.
\label{Lsmall}
\end{eqnarray}
The solid curve in Fig.\ \ref{Deltagammaedgeplot} shows $\Delta \gamma (W/2)$ normalized to $B W^2/ \phi_0$ as a function of $L/W$ along with the limiting behaviors of Eqs.\ (\ref{Lbig}) (dot-dashed) and (\ref{Lsmall}) (dotted).  The dashed curve shows the interpolating function, 
\begin{equation}
\Delta \gamma(W/2) = \frac{14 \zeta(3) B W^2}{\pi^2 \phi_0}
\tanh\Big[\frac{\pi^3 L}{28 \zeta(3) W}\Big],
\label{edgeapprox}
\end{equation}
where $\zeta(3) = 1.20206$ is the Riemann zeta function.

\begin{figure}
\includegraphics[width=8cm]{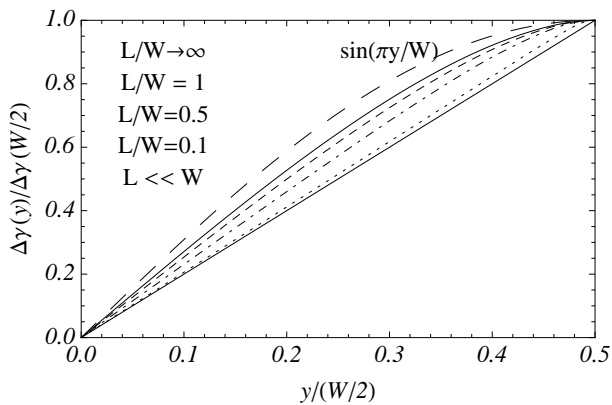}
\caption{%
$\Delta \gamma (y)/\Delta \gamma (W/2)$ vs $y/(W/2)$ for $L \to \infty$ (upper solid curve), $L/W = 1$ (dashed),  $L/W = 0.5$ (dot-dashed), $L/W = 0.1$ (dotted), and   $L \ll W$ (lower solid curve), for which $\Delta \gamma (y)/\Delta \gamma (W/2) = y/(W/2)$.  For comparison, the top long-dashed curve shows $\sin (\pi y/W)$.}
\label{Deltagammanormplot}
\end{figure} 

The plots of  $\Delta \gamma (y)/\Delta \gamma (W/2)$ vs $y/(W/2)$ in Fig.\ \ref{Deltagammanormplot} show how the gauge-invariant phase difference depends upon the ratio $L/W$.  For $L/W \to \infty$, the curve lies below $\sin (\pi y/W)$, shown as the long-dashed curve, and for $L \ll W$,  $\Delta \gamma (y)/\Delta \gamma (W/2) = y/(W/2)$, a straight line.

\begin{figure}
\includegraphics[width=8cm]{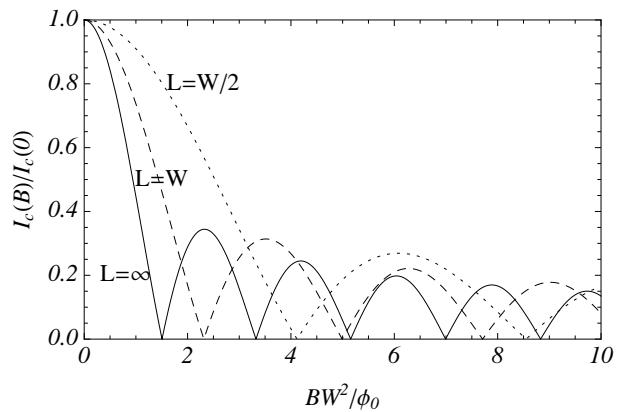}
\caption{%
$I_c(B)/I_c(0)$ vs $BW^2/\phi_0$ calculated from Eqs.\ (\ref{Deltagamma}) and (\ref{IcBnorm}) for $L/W = \infty$ (solid), 1 (dashed) and 1/2 (dotted). }
\label{3Ls}
\end{figure}

The maximum Josephson current $I_c(B)$, the maximum integral of $j_c d \sin \Delta \gamma(y)$ over $y$ from $-W/2$ to $W/2$, occurs when $\Delta \gamma_0 =\pm \pi/2$, such that 
\begin{equation}
\frac{I_c(B)}{I_c(0)} = \frac{2}{W}\Big|\int_0^{W/2} 
\cos [\Delta \gamma(y)]dy \Big|,
\label{IcBnorm}
\end{equation}
where $\Delta \gamma(y)$ is given in Eq.\ (\ref{Deltagamma}).  
Figure \ref{3Ls} shows plots of $I_c(B)/I_c(0)$ vs $B W^2/\phi_0$ for $L/W = \infty,$ 1, and 1/2.  The stretching out of the pattern along the horizontal axis as $L$ decreases is easily understood with the help of Fig.\ \ref{Deltagammaedgeplot}.

Let us define $\Delta B_1$ as the value of $B$ at which $I_c(B)$ has its first zero, $\Delta B_2$ as the difference of the values at which $I_c(B)$ has its second and first zeros, and $\Delta B_n$ as the difference of the values at which $I_c(B)$ has its $n$th and  $(n-1)$th zeros. 
For all finite values of $L/W$, the $\Delta B_n$ are smaller for small $n$ than for large $n$. However, for large $n$, the $\Delta B_n$ approach the limiting value
\begin{equation}
\Delta B =\Big(\frac{\pi^3 \phi_0}{16 W^2}\Big)/\sum_{n=0}^\infty 
\frac{\tanh[(2n+1)\pi L/4W]}{(2n+1)^3}.
\label{DeltaB}
\end{equation}
To illustrate this, if we approximate $\Delta \gamma (y)$ for $L \to \infty$ by  $\Delta \gamma(W/2) \sin(\pi y/W)$ (see the long-dashed curve in Fig.\ \ref{Deltagammanormplot}), as in Ref.\ \onlinecite{Moshe08}, then the integral in Eq.\ \ref{IcBnorm} can be evaluated in terms of the Bessel function $J_0$ with the result 
\begin{equation}
\frac{I_c(B)}{I_c(0)} = \Big|J_0\big(\frac{14 \zeta(3) B W^2}{\pi^2 \phi_0}\big)\Big|.
\label{IcBLbig}
\end{equation}
For $L/W \to \infty$, the sum in Eq.\ (\ref{DeltaB}) can be evaluated as 
\begin{equation}
\Delta B = [\pi^3/14 \zeta (3)]\phi_0/W^2= 1.842\phi_0/W^2,
\label{DeltaBlargeL}
\end{equation}
as pointed out in Ref.\ \onlinecite{Rosenthal91}. 
Using Eq.\ (\ref{IcBLbig}) and the well-known zeros of $J_0(x)$, we find the following values for $n$ = 1, 2, 3, 4, and 5:  $\Delta B_n/\Delta B$ = 0.7655, 0.9916,  0.9975, 0.9988, and 0.9993.  However, to evaluate the zeros of $Ic(B)/I_c(0)$ without using the Bessel-function approximation, we must numerically evaluate Eq.\ (\ref{IcBnorm}).  This yields   the following more accurate values for $n$ = 1, 2, 3, 4, and 5:  $\Delta B_n/\Delta B$ = 0.8173, 0.9866,  0.9946, 0.9968, and 0.9979.

When $d_i \ll L \ll W$, the gauge-invariant phase difference $\Delta \gamma (y)$ of Eq.\ (\ref{Deltagamma}) becomes linear in $y$, and in this case we have the familiar Fraunhofer-like pattern, 
\begin{equation}
\frac{I_c(B)}{I_c(0)} = \Big|\frac{\sin(\pi BWL/2\phi_0)}{\pi BWL/2\phi_0}\Big|,
\label{IcBLsmall}
\end{equation}
such that all the $\Delta B_n$ are the same and equal to 
\begin{equation}
\Delta B = 2 \phi_0/WL.
\label{DeltaBsmallL}
\end{equation}
The magnitude of this $\Delta B$ agrees with that in sandwich-type Josephson junctions with thickness $d \gg \lambda$ along the $z$ direction only in the limit  $d_i \ll L < \lambda$.\cite{Weihnacht69,Barone82,Orlando91}

\begin{figure}
\includegraphics[width=8cm]{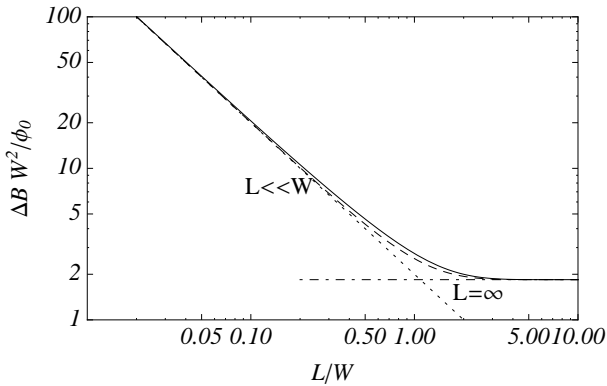}
\caption{%
$\Delta B$ vs $L/W$ calculated from Eqs.\ (\ref{DeltaB}) (solid), (\ref{DeltaBlargeL}) (dot-dashed), (\ref{DeltaBsmallL}) (dotted), and (\ref{DeltaBinterpolating}) (dashed). }
\label{DeltaBvsL}
\end{figure}

The solid curve in Fig.\ \ref{DeltaBvsL} shows $\Delta B$, the large-$n$ limit of $\Delta B_n$, calculated via Eq.\ (\ref{DeltaB}) as a function of $L/W$, along with the expressions for $\Delta B$ in the limits $L/W \to \infty$ (dot-dashed)  and $L \ll W$ (dotted).  The dashed curve shows the approximate interpolating function obtained from Eq.\ (\ref{edgeapprox}),
\begin{equation}
\Delta B  = \Big(\frac{\pi^3 \phi_0}{14 \zeta(3) W^2}\Big)/
\tanh\Big[\frac{\pi^3 L}{28 \zeta(3) W}\Big].
\label{DeltaBinterpolating}
\end{equation}

\section{\label{summary}Summary}

In this paper I have considered a Josephson junction bisecting a rectangular  superconducting  thin film of large Pearl length $\Lambda = 2 \lambda^2/d$ subjected to a perpendicular magnetic induction $B$.  I calculated the gauge-invariant phase difference and used it to determine the $B$ dependence of the Josephson critical current density $I_c(B)$.

\begin{acknowledgments}
I thank J. E. Sadleir and V. G. Kogan for stimulating discussions.
My work at the Ames Laboratory was
supported by the Department of Energy - Basic Energy Sciences under Contract
No. DE-AC02-07CH11358.

\end{acknowledgments}

\appendix 

\section{\label{infL}%
The limit $L \to \infty$} 

In the limit $L \to \infty$, the gauge-invariant phase difference $\Delta \gamma (y)$ given in Eq.\ (\ref{Deltagamma}) can be expressed as 
$\Delta \gamma (y) =(16 B W^2/\pi^2 \phi_0) \sigma(\pi y/W)$,
where
\begin{eqnarray}
\sigma(\psi)&=&\sum_{n=0}^\infty (-1)^n \sin[(2n+1)\psi]/(2n+1)^3 \\
&=&(i/16)e^{-i\psi}[\Phi(-e^{-2i\psi},3,1/2) \nonumber \\
&&-e^{2i\psi}\Phi(-e^{2i\psi},3,1/2)],
\end{eqnarray}
and $\Phi(z,s,a)=\sum_{k=0}^\infty z^k/(k+a)^s$
is the Lerch transcendent.\cite{Mathematica}
Note that $\sigma(0)=0$ and $\sigma(\pi/2) = 7 \zeta(3)/8 = 1.0518.$


\begin{thebibliography}{99}
\bibitem{VanHarlingen95} D. J. Van Harlingen, \rmp {bf 67}, 515 (1995).
\bibitem{Delin96} K. A. Delin and A. W. Kleinsasser, Supercond. Sci. Technol. {\bf 9}, 227 (1996).
\bibitem{Hilgenkamp02} H. Hilgenkamp and J. Mannhart, \rmp, {\bf 74}, 485 (2002).
\bibitem{Durrell09} J. H. Durrell and N. A. Rutter, Supercond. Sci. Technol. {\bf 22}, 013001 (2009).
\bibitem{Mints94} R. G. Mints and I. B. Snapiro, \prb {\bf 49}, 6188 (1994).
\bibitem{Kogan01} 	V. G. Kogan, V. V. Dobrovitski, J. R. Clem, Y. Mawatari, and R. G. Mints, \prb {\bf 63}, 144501 (2001) 
\bibitem{Moshe08} M. Moshe, V. G. Kogan, and R. G. Mints, \prb {\bf 78}, 020510(R) (2008).
\bibitem{Pearl64} J. Pearl, \apl {\bf 5}, 65 (1964).
\bibitem{Ivanchenko90} Yu. M. Ivanchenko and T. K. Soboleva, Phys. Lett. A {bf 147}, 65 (1990).
\bibitem{Gurevich92} A. Gurevich, \prb {\bf 46}, 3187 (1992).
\bibitem{Ivanchenko95} Yu. M. Ivanchenko, \prb {\bf 52}, 79 (1995).
\bibitem{Kuzovlev97} Yu. E. Kuzovlev and A. I. Lomtev, Sov. Phys. JETP {\bf 84}, 986 (1997).
\bibitem{deGennes} P. G. de Gennes, {\it Superconductivity of Metals and Alloys} (Benjamin, New York, 1966), p. 177.
\bibitem{StJames69} D. Saint-James, E. J. Thomas, and G. Sarma, {\it Type II Superconductivity} (Pergamon, Oxford, 1969).
\bibitem{Rosenthal91} P. A. Rosenthal, M. R. Beasley, K. Char, M. S. Colclough, and G. Zaharchuk, \apl {\bf 59}, 3482 (1991).
\bibitem{Weihnacht69} M. Weihnacht, Phys. Stat. Sol. {\bf 32}, K169 (1969).
\bibitem{Barone82} A. Barone and G. Paterno, {\it Physics and Applications of the Josephson Effect}, (Wiley, New York, 1982).
\bibitem{Orlando91} T. P. Orlando and K. A. Delin, {\it Foundations of Applied Superconductivity}, (Addison-Wesley, Reading, 1991), p. 447.
\bibitem{Mathematica} Wolfram Research, Inc., Mathematica, Version 7.0, Champaign, IL (2008).
\end{thebibliography}
\end{document}